# Probing the emission physics and weak/soft population of Gamma-Ray Bursts with *LOFT*

## White Paper in Support of the Mission Concept of the Large Observatory for X-ray Timing

Authors

L. Amati[1], G. Stratta[2], J-L. Atteia[3], M. De Pasquale[4], E. Del Monte[5], B. Gendre[6], D. Götz[7], C. Guidorzi[8], L. Izzo[9], C. Kouveliotou[10], J. Osborne[11], A.V. Penacchioni[9], P. Romano[4], T. Sakamoto[12], R. Salvaterra[13], S. Schanne[7], J. J. M. in 't Zand[14], L.A. Antonelli[5], J. Braga[15], S. Brandt[16], N. Bucciantini[17], A. Castro-Tirado[18], V. D'Elia[19], M. Feroci[5], F. Fuschino[5], D. Guetta[20], F. Longo[21], M. Lyutikov[22], T. Maccarone[23], V. Mangano[24], M. Marisaldi[1], S. Mereghetti[13], P. O'Brien[11], E.M. Rossi[25], F. Ryde[26], P. Soffitta[5], E. Troja[10], R.A.M.J. Wijers[27], B. Zhang[28]

[1] IASF Bologna, Italy
[2] University of Urbino, Urbino, Italy
[3] IRAP Toulouse, France
[4] IASF Palermo, Italy
[5] IASF Roma, Italy
[6] ARTEMIS, Nice, France
[7] IRFU/CEA Saclay, 91191 Gif sur Yvette, France
[8] University of Ferrara, Ferrara, Italy
[9] ICRANet, Italy
[10] NASA Marshall Space Flight Center, Huntsville, AL, USA
[11] University of Leicester, Leicester, Italy
[12] Aoyama Gakuin University, Tokyo, Japan
[13] IASF Milano, Italy
[14] SRON, Utrecht, The Netherlands
[15] INPE, Sao José dos Campos, Brazil
[16] Technical University of Denmark, Lyngby, Denmark
[17] Osservatorio Astrofisico di Arcetri, Firenze, Italy
[18] IAA, Spain
[19] ASI Science Data Center, Italy
[20] Observatory of Rome, Rome, Italy
[21] University of Trieste, Trieste, Italy
[22] Purdue University, West Lafayette, IN, USA
[23] Texas Tech University, Lubbock, TX, USA
[24] Penn State University, PA, USA
[25] University of Leiden, Leiden, Netherlands
[26] University of Stockholm, Stockholm, Sweden
[27] University of Amsterdam, Amsterdam, Netherlands
[28] University of Nevada, Las Vegas, NV, USA

## Preamble

The Large Observatory for X-ray Timing, *LOFT*, is designed to perform fast X-ray timing and spectroscopy with uniquely large throughput (Feroci et al., 2014a). *LOFT* focuses on two fundamental questions of ESA's Cosmic Vision Theme "Matter under extreme conditions": what is the equation of state of ultra-dense matter in neutron stars? Does matter orbiting close to the event horizon follow the predictions of general relativity? These goals are elaborated in the mission Yellow Book (`http://sci.esa.int/loft/53447-loft-yellow-book/`) describing the *LOFT* mission as proposed in M3, which closely resembles the *LOFT* mission now being proposed for M4.

The extensive assessment study of *LOFT* as ESA's M3 mission candidate demonstrates the high level of maturity and the technical feasibility of the mission, as well as the scientific importance of its unique core science goals. For this reason, the *LOFT* development has been continued, aiming at the new M4 launch opportunity, for which the M3 science goals have been confirmed. The unprecedentedly large effective area, large grasp, and spectroscopic capabilities of *LOFT*'s instruments make the mission capable of state-of-the-art science not only for its core science case, but also for many other open questions in astrophysics.

*LOFT*'s primary instrument is the Large Area Detector (LAD), a 8.5 $m^2$ instrument operating in the 2–30 keV energy range, which will revolutionise studies of Galactic and extragalactic X-ray sources down to their fundamental time scales. The mission also features a Wide Field Monitor (WFM), which in the 2–50 keV range simultaneously observes more than a third of the sky at any time, detecting objects down to mCrab fluxes and providing data with excellent timing and spectral resolution. Additionally, the mission is equipped with an on-board alert system for the detection and rapid broadcasting to the ground of celestial bright and fast outbursts of X-rays (particularly, Gamma-ray Bursts).

This paper is one of twelve White Papers that illustrate the unique potential of *LOFT* as an X-ray observatory in a variety of astrophysical fields in addition to the core science.

## 1 Summary

The Large Observatory For X-ray Timing (*LOFT*, Feroci et al. 2012, 2014b) is a space mission concept studied by ESA during an extended assessment phase in 2011–2013 as a candidate for an M3 mission within the Cosmic Vision Programme, and proposed as a M4 mission. *LOFT* aims at probing gravity in the strong field environment of black holes and other compact objects, and investigating the state of matter at supra-nuclear densities in neutron stars through X-ray timing with unprecedented sensitivity. The payload is composed of the Large Area Detector (LAD), in the 2–30 keV energy band, a peak effective area of about 8.5 $m^2$ and an energy resolution better than 260 eV, and the Wide Field Monitor (WFM), a coded mask imager with a field of view of 4.1 steradians, an energy resolution of about 300 eV and a point source location accuracy of 1 arcmin in the 2–50 keV energy range (see Table 1). The WFM (Brandt et al., 2012) will monitor approximately 1/3 of the sky at any time.

As detailed in the next sections, thanks to its unique combination of field of view (FOV), energy band, energy resolution and source location accuracy, combined with the capability of promptly transmitting Gamma Ray Burst (GRB) positions to the ground, the *LOFT*/WFM will achieve scientific goals of fundamental importance and not fulfilled by GRB experiments presently flying (e.g., *Swift*, *Konus*/WIND, *Fermi*/GBM, MAXI) and future approved missions (see Table 2). These goals can be summarized as follows:

- substantial increase (with respect to the past and current missions) in the detection rate of X-Ray Flashes (XRF), a sub-class of soft events which constitutes the bulk of the GRB population and still lacking a good body of observational data. With respect to HETE-2/WXM (2–25 keV), the most efficient XRF detector flown so far, *LOFT*/WFM will increase the XRF detection rate 7-fold, with an expected rate of 30–40 XRF per year;

- measurement of the GRB spectra and their evolution down to 2 keV as an indispensable tool for testing models of GRB prompt emission;

- detection and study of the rarely detected transient X-ray absorption features in medium/bright GRBs, particularly during their initial stage. These measurements are of paramount importance for the understanding of the properties of the Circum-Burst Matter (CBM) and hence the nature of GRB progenitors, which is still a major open issue in the field. The detection of transient features can enable us to determine the GRB redshift;

- extension of GRB detection up to very high redshift ($z > 6$), which is of fundamental importance for the study of luminosity evolutionary effects, the measurement of the star formation rate, the evolution of the properties of the interstellar medium, and possible for the discovery of population III stars;

- fast and accurate localization of GRBs to allow prompt multi-wavelength follow-up with ground and space telescopes. This will in turn lead to the identification of the optical counterparts and/or host galaxies and the redshift. GRB cosmological redshift is a fundamental measurement for the scientific goals listed above.

## 2 Introduction

Discovered in the late 1960s by military satellites and revealed to the scientific community in 1973, Gamma-ray Bursts (GRBs) are one of the most intriguing "mysteries" of modern science (Mészáros, 2006; Gehrels et al., 2009). Although being very bright (with fluences up to more than $10^{-4}$ erg cm$^{-2}$ released in a few tens of s) and very frequent (about 0.8/day as measured by low earth orbit satellites), their origin and physics remained obscure for more than 20 years. Huge observational efforts in the last decades have provided, among others, 1) a good characterization of the burst-temporal and spectral properties, 2) the accurate localization and discovery of their multi-wavelength afterglow emission by *BeppoSAX* and *Swift*, 3) the determination of their cosmological distance scale, and 4) the evidence of a connection of long GRBs (i.e. GRBs with burst duration $> 2$ s, Kouveliotou et al. 1993) with type Ib/c supenovae (SNe). However, our understanding of the GRB phenomenon is still incomplete due to several open issues, both observational and theoretical. These include: identification and understanding of sub-classes of GRBs (short/long, X-ray Flashes, sub-energetic), physics and geometry of the prompt emission, unexpected early afterglow phenomenology (e.g. plateau and flares), the GRB/SN connection, the cosmological use of GRBs, VHE emission, nature of the inner engine, and more.

Table 1: Main characteristics of the *LOFT* instruments (based on Feroci et al., 2014b; Zane et al., 2014; Brandt et al., 2014).

| | LAD | WFM |
|---|---|---|
| Energy band (keV) | 2–30(80) | 2–50 |
| Effective area | 8.5 $m^2$ (at 6 keV) | 90 $cm^2$ (peak) |
| FOV | 1 deg | 4.1 sr |
| Sensitivity (5$\sigma$) | 0.1 mCrab in 100s | 0.6 Crab in 1 s<br>2.1 mCrab in 50 ks |
| Energy resolution | 180–240 eV | 300 eV |
| Timing resolution | 10 $\mu$s | 10 $\mu$s |
| Source location | – | 0.5 – 1 arcmin |

Table 2: Characteristics of the *LOFT*/WFM compared with the past and present experiments with major impact in GRB science, as well with future GRB experiments under development or advanced study: the French-Chinese mission SVOM (Godet et al., 2012), Lomonosov/UFFO–p, an experiment by an international collaboration led by Korea and Taiwan to be launched on board the Russian Lomonosov mission (Kim et al., 2012), the GRB Monitor of the Japanese CALET experiment on the International Space Station (ISS) (Yamaoka et al., 2013). See Sect. 3.

| | Energy Band | FOV sr | Energy resolution keV | Peak eff. area $cm^2$ | Source location accuracy | Operation |
|---|---|---|---|---|---|---|
| CGRO/BATSE | 20 keV–2 MeV | $4\pi$ | 10 (100) | $\sim$ 1700 | $> 1.7°$ | ended |
| BeppoSAX/WFC | 2–28 keV | 0.25 | 1.2 (6) | 140 | $1'$ | ended |
| HETE-2/WXM | 2–25 keV | 0.8 | 1.7 (6) | 350 | 1–3′ | ended |
| Swift/BAT | 15–150 keV | 1.4 | 7 (60) | ~2000 | 1–4′ | active |
| Fermi/GBM | 8 keV – 40 MeV | $4\pi$ | 10 (100) | 126 | $> 3°$ | active |
| Konus–WIND | 20 keV – 15 MeV | $4\pi$ | 10 at 100 keV | 120 | – | active |
| SVOM | 4 keV – 5 MeV | 2 | 2 (60) | 400 | 2–10′ | 2021-2025 |
| Lomonosov/UFFO–p | 5–100 keV | 1.5 | 2 (60) | 191 | 5–10′ | 2015–2020? |
| CALET/GBM | 7 keV – 20 MeV | 3 | 5 (60) | 68 | – | 2015–2018? |
| *LOFT*/WFM | 2–50 keV | 4.1 | 0.3 (6) | 90 | 0.5–1′ | 2025–2028 |

In this document we briefly review in Sect. 3 the main GRB scientific topics that *LOFT*/WFM can accomplish. In Sect. 4 we discuss X-ray afterglow follow-up with the LAD and GRB observations through the LAD collimator. The *LOFT* Burst Alert System and the role that *LOFT* could play in the multiwavelength and multimessenger astronomy is described in Sect. 5.

## 3 GRB Science with the WFM

Many issues are still open on GRBs. In this section we briefly describe the most relevant GRB science that can be addressed with the *LOFT*/WFM observations.

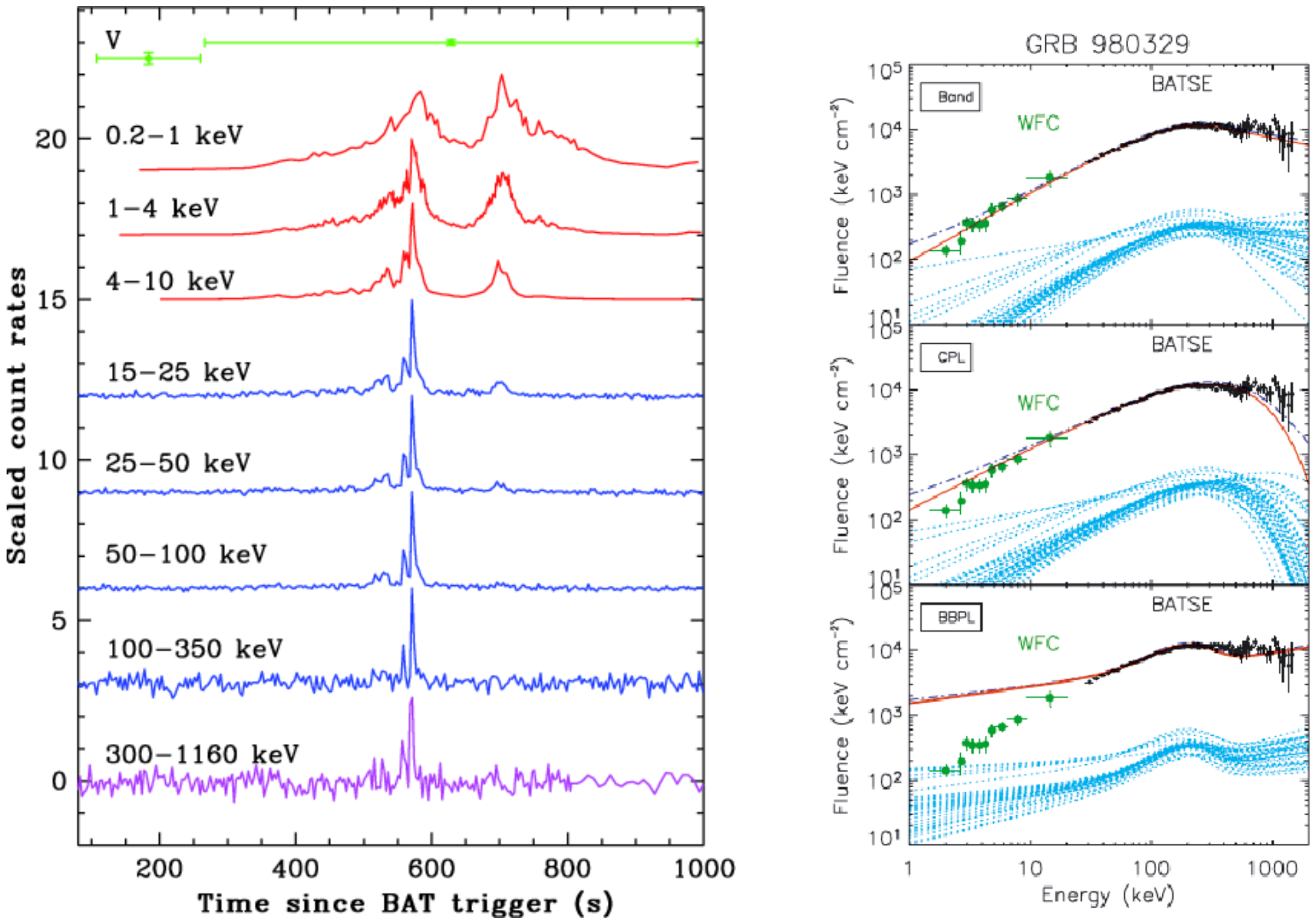

Figure 1: Examples of measurements with 1–1000 keV coverage. *Left: Swift* UVOT, XRT, BAT and Konus light curves of GRB 060124. The count rates have been normalized to the peak of each light curve and offset vertically for clarity (Romano et al., 2006). *Right:* Extrapolation to X-rays of the best fit models (Band function, cut-off power law and BB + power-law) of the BATSE spectrum of GRB 980329 (Ghirlanda et al., 2007). A persistent "quasi thermal" model could be ruled out based on low energy (2–28 keV) WFC data.

## 3.1 Physics of the prompt emission

Long GRBs are most probably connected to the collapse of massive fast rotating stars (collapsars, e.g., Woosley, 1993; Paczyński, 1998). In the most extensively discussed scenario, the collapsar model, a massive star explodes as a supernova (SN) and the collapsing core produces a black hole: the GRB originates from a jet emerging along the rotation axis. The basic properties of the GRB afterglow emission (flux and spectral shape) can be satisfactorily explained in the framework of the "standard" fireball model plus the external shock scenario (Rees & Meszaros, 1992; Sari et al., 1998), where shocks are produced by the interaction of the GRB collimated, relativistic ejecta with the surrounding interstellar matter. On the contrary, the physics underlying the complex light curves and the fast spectral evolution of the "prompt" emission (i.e., the burst itself) is still far from understood. Presently, there is a forest of models invoking different origins of fireball (kinetic energy dominated, Poynting flux driven), of shocks (internal, external) and of emission mechanisms (synchrotron and/or Inverse Compton originated in the shocks, direct or Comptonized thermal emission from the fireball photosphere, and mixtures of these). These models could not be disentangled with recent Fermi observations (Zhang, 2014).

A simple analysis of the morphology of the light curve as a function of energy shows that a wealth of information is contained in the prompt X-ray emission of GRBs. For instance, the broadening and increasing time lag of the pulses towards lower photon energies is also a crucial test for the emission models (Fig. 1, left). Past observations with BeppoSAX/WFC (2-28 keV) have demostrated that the measurements of the low energy (<10 keV) light curves and spectra are of key importance to discriminate among different emission models. An important circumstance is that for a large fraction of GRBs, the slope of the time integrated prompt X-ray (<10 keV) emission spectrum is inconsistent with the predictions of optically thin synchrotron shock models (Preece et al., 2000; Frontera & et al., 2000). Inconsistency was found also with Comptonized thermal emission of the photosphere (Ghirlanda et al., 2007) as shown in Fig. 1 (right), pointing out that, if present, this component should rapidly evolve with time (see also e.g. Frontera et al. 2013).

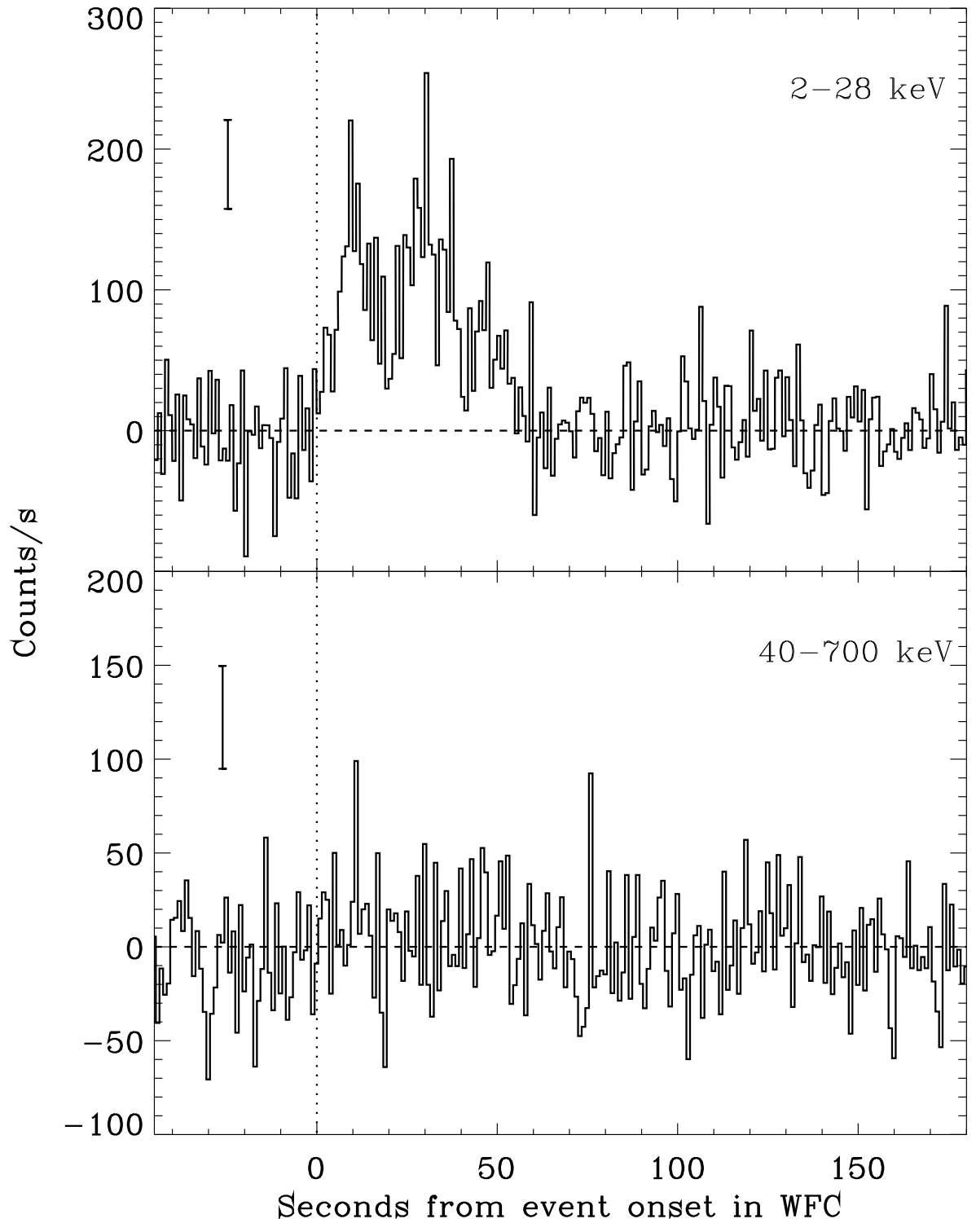

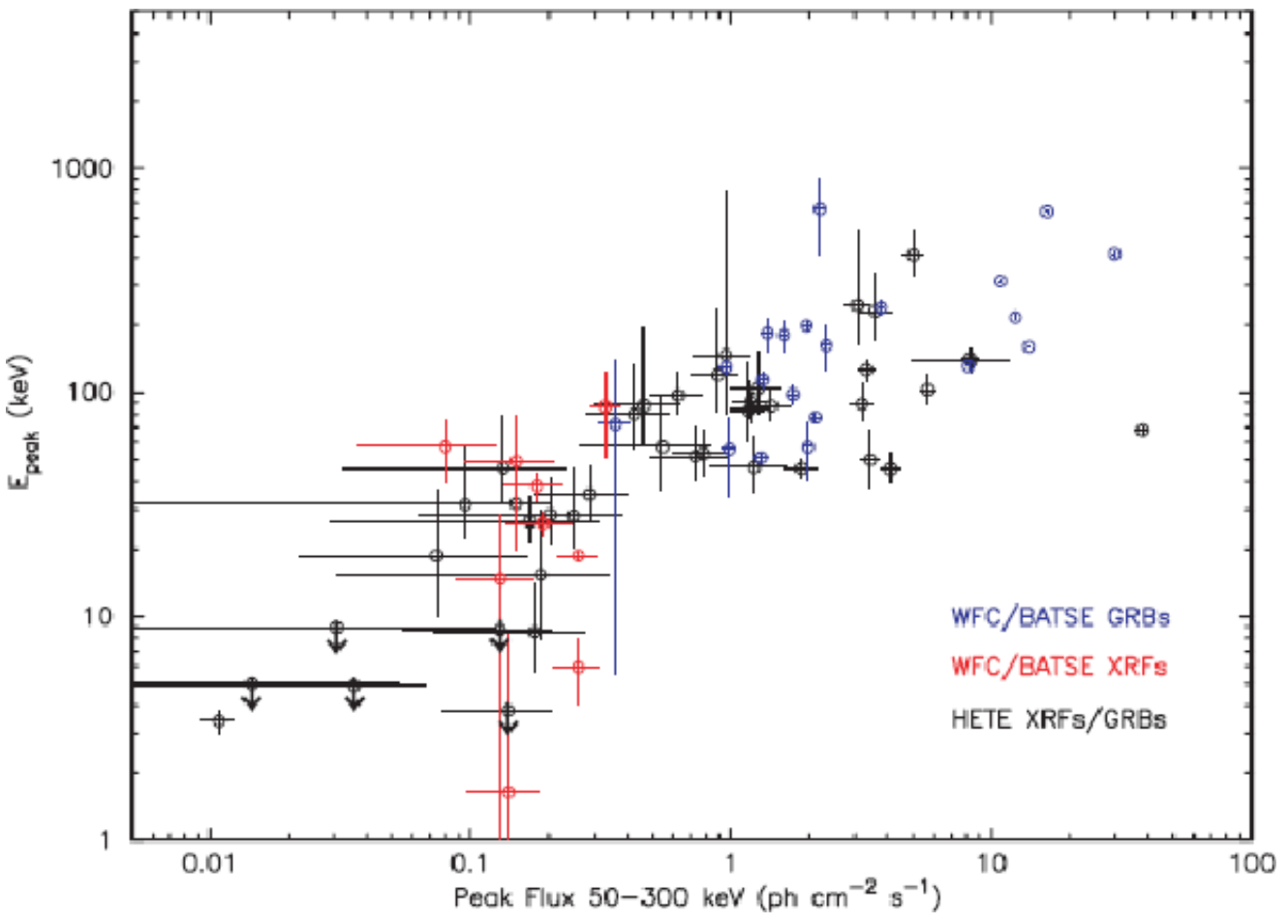

Figure 2: X-ray flashes will be one of the GRB science main targets of *LOFT*/WFM. *Left:* 1 s background subtracted light curve of the X-ray Flash XRF 020427 in the BeppoSAX/WFC 2–28 keV energy band (Amati et al., 2004). *Top:* Spectral peak energy $E_{peak}$ as a function of peak flux, showing the existence of a sub-class of weak GRBs with very low values of $E_{peak}$ (i.e., XRF events) which can be sensitively investigated only by detectors with energy threshold down to 1–2 keV.

Another intriguing case is the detection by BeppoSAX WFCs plus GRBM of a transient emission feature in the prompt emission of GRB990712 (Frontera et al., 2001) that was interpreted as evidence of blackbody emission from the fireball photosphere. Evidence of an evolving thermal black body spectrum (with typical temperatures of 5 keV) was discovered also in the long GRB 060218, for which Swift/XRT observations during the prompt emission were available (Campana et al., 2006), and interpreted as due to the supernova shock breaking out of the stellar envelope.

## 3.2 X-ray flashes

Since ~2000, a new class of fast transients (X-Ray Flashes, XRFs) has been discovered and studied, first by *BeppoSAX* (Heise et al., 2001) and then by *HETE-2* (Sakamoto et al., 2005; Pélangeon et al., 2008). XRFs show similar durations as long GRBs, are isotropically distributed in the sky, have a similar rate of occurrence and are associated with very similar type Ib/c SNe that are found to accompany standard long GRBs (Soderberg et al., 2005). Their main property is however that most of the prompt emission occurs in the X-ray band (2–20 keV), with negligible emission at higher energy (Fig. 2). An empirical definition of an XRF was given as a GRB with a 2–30 keV over 30-400 keV flux ratio greater than unity (Sakamoto et al., 2004).

As shown by Kippen et al. (2003), XRFs fall naturally in the low energy wing of the GRB peak energy distribution. This class is likely a lower energy extension of GRBs. XRFs satisfy the ‘Amati’ relation between redshift-corrected peak energy $E_p$ in the $\nu F_\nu$ prompt emission spectrum and released energy $E_{iso}$ (Amati et al., 2002; Amati, 2006). Analysis based on the complete HETE-2 spectral catalog (Pélangeon et al., 2008) showed that most XRFs lie at low/moderate redshift and, importantly, they are likely predominant in the GRB population and closely linked to the 'classical' GRBs (Fig. 2, right).

In the fireball model paradigm for GRBs, XRFs could be characterized by a lower bulk Lorenz factor than GRBs, possibly due to a high baryon load of the fireball ('dirty fireball'). A 'clean fireball' may also produce XRFs, because a gas outflow with higher Lorentz factor would cause internal shocks at larger radii where magnetic field density (or equivalently internal energy) is smaller (Mochkovitch et al., 2004). XRFs may be the missing link between the more numerous population of hard/luminous GRBs and the class of relatively soft/low luminosity GRBs associated with a SN event. In addition, very soft XRFs (expected detection rate ~30–50 $yr^{-1}$) may form a sub-class of soft/ultrasoft events which may constitute the bulk of the GRB population (Pélangeon et al., 2008). It is thus important to extend the XRF sample to confirm these predictions. This can be done only by extending the band down to soft X-rays.

### 3.3 X-ray spectral features

Several long GRBs were associated with supenova emission (Stanek et al., 2003). However there are a few cases with no SN. In addition, there are cases in which SNe with extreme expansion velocity of their remnants, that is the main characteristic of GRB-SN, have no simultaneous GRB events associated with them, as in the cases of SN2002ap (Wang et al., 2003), 2009bb (Soderberg et al., 2010), and that of 2012ap (Chakraborti et al., 2014). Several GRB progenitor scenarios are debated (e.g. collapsar, supranova, induced gravitational collapse). Probing the chemical abundance in the circumburst environment of GRBs is a possible way to test progenitor models (Levesque et al., 2010). This task is extremely challenging in the optical, however, and X-ray spectroscopy of the early phases of the GRB prompt emission provides a complementary, and, in many respects, a unique tool for this goal.

Past observations with *BeppoSAX* showed variable absorption features in the X-ray spectrum of the prompt emission of a few GRBs, namely GRB000528 (Frontera et al., 2004), GRB010214 (Guidorzi et al., 2003), GRB990705 (Amati et al., 2000). These features were present in the first part (rise) of the burst and faded away soon after. This is in line with theoretical expectations because most of the opacity evolution, which is principally due to photoionization of gas-phase ions and of dust grain destruction, takes place in the early phase of the event (Böttcher et al., 1999; Lazzati & Perna, 2003). The observed features point to the presence of an iron-rich circumburst environment, but no conclusive constraints on the progenitor model were achieved given the low statistical quality. The *Swift* mission (Gehrels et al., 2004), in spite of the enormous effort to promptly follow-up (slew time ~ 1 min) in the 0.3–10 keV band the GRBs detected with *Swift*/BAT in the hard X-rays (15-150 keV), has been unable to study the early phases of the prompt emission down to 1 keV, when the absorption features are expected (e.g., in the case of GRB990705, the absorption feature was visible only in the first 13 s).

### 3.4 High-$z$ GRBs

Our observational window on the Universe extends up to $z \sim 10$, where the most distant objects discovered so far reside, and then recovers at $z = 1000$, the epoch of primordial fluctuations measured by *BOOMERANG*, *WMAP*, and *Planck*. The formation of the first objects, stars, and protogalaxies, should have taken place at epochs corresponding to $z = 10$–$30$, certainly beyond $z = 7$. Discovering and studying the first "light" from primordial gravitationally bounded objects in the Universe at such redshifts is thus a primary goal of cosmology and astrophysics. Far infrared and X-ray radiation are the only two wavelength regimes in which these studies can be attempted, with the X-ray revealing the most energetic part of the phenomenon.

The increase of the number of high redshift GRBs and their few-arcmin localization, needed to allow optical follow-up and hence redshift determination, is of fundamental importance not only for GRB physics and progenitors, but also for the study of the star formation rate evolution, of population III stars and, more in general, of the properties of the universe at the ionization epoch. Several recent studies (Salvaterra et al., 2008, 2009) show that lowering the threshold of the GRB detector energy band can increase substantially the sensitivity to high-$z$ GRB (see also Fig.3, right).

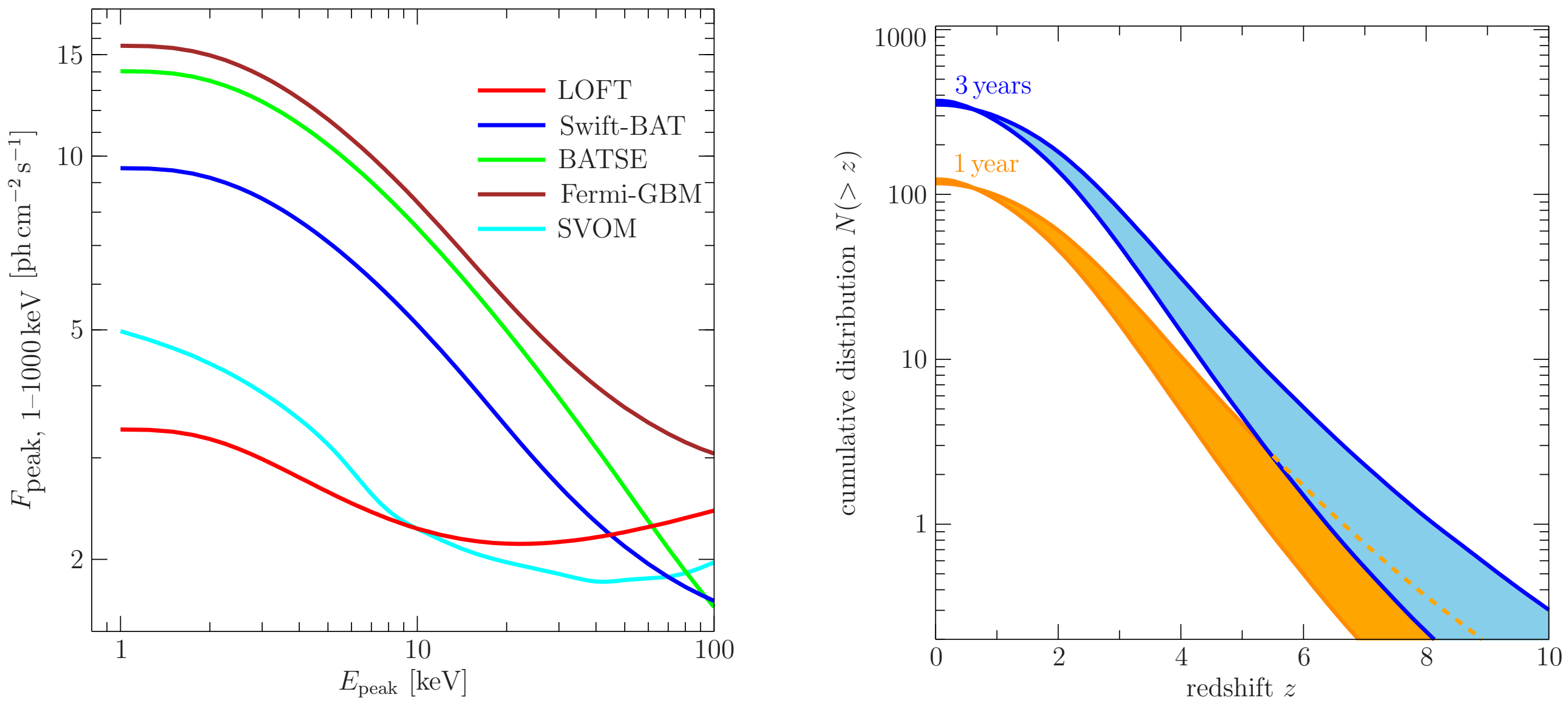

Figure 3: Illustrative plots of *LOFT*/WFM capabilities and of their impact in GRB science. *Left:* GRB detection sensitivity in terms of peak flux sensitivity as a function of the spectral peak energy $E_p$ (Band, 2003) of the *LOFT*/WFM (red) for M4 configuration compared to those of *CGRO*/BATSE (green), *Swift*/BAT (blue), *Fermi*/GBM (brown), and *SVOM*/ECLAIRs (cyan). *Right:* Cumulative redshift distribution of long GRBs predicted for *LOFT*/WFM and computed assuming a broken power-law GRB luminosity function and a pure density evolution (see Salvaterra et al., 2012, for the details of the model). At the flux limit of *LOFT*/WFM we predict about 1–2 GRBs yr$^{-1}$ at $z > 6$, corresponding to 3–6 high-$z$ GRBs in 3 years.

## 3.5 SN shock breakout

Wide-field soft X-ray surveys are predicted to detect each year hundreds of supernovae in the act of exploding through the SN shock breakout phenomenology. According to recent SN rate estimates (Horiuchi et al., 2011), about two SNe breakouts per year should be detected with the *LOFT* wide field monitor within 20 Mpc.

SN shock breakout events occur in X-rays at the very beginning of the supernova explosions, allowing for follow-up at other wavelengths to start extremely early on in the supernova, and possibly shedding light on the nature of the progenitor of some types of SNe. To date, only one SN shock breakout has been unambiguously detected, in *Swift* observations of NGC 2770. It has a fast-rise, exponential decay light curve with a rise time of 72 seconds, and an exponential decay timescale of 129 seconds, and a peak X-ray luminosity of $6.1 \times 10^{43}$ ergs s$^{-1}$ (Soderberg et al., 2008). More uncertain is the case of GRB 060218, the X-ray prompt emission of which (observed with Swift/XRT) was interpreted as the evidence of a SN shock breakout (Campana et al., 2006).

If we take the case of NGC 2770 as a template, the optimum integration time to detect such events out to 20 Mpc is 240 s (i.e., WFM sensitivity ~30 mCrab).

## 3.6 Expected performance of the *LOFT*/WFM

*LOFT*/WFM addresses most of the GRB scientific open issues discussed in the previous section. The WFM outperforms in spectral resolution, source location accuracy, and field of view the two earlier main instruments capable of measuring GRB prompt emission down to 2 keV (*BeppoSAX*/WFC and *HETE-2*/WXM) as well as other past, present and future high energy instruments (Table 2). This will allow WFM, possibly in combination with other GRB experiments flying at the same epoch, to provide a unique contribution to the fulfilment of the scientific goals mentioned above.

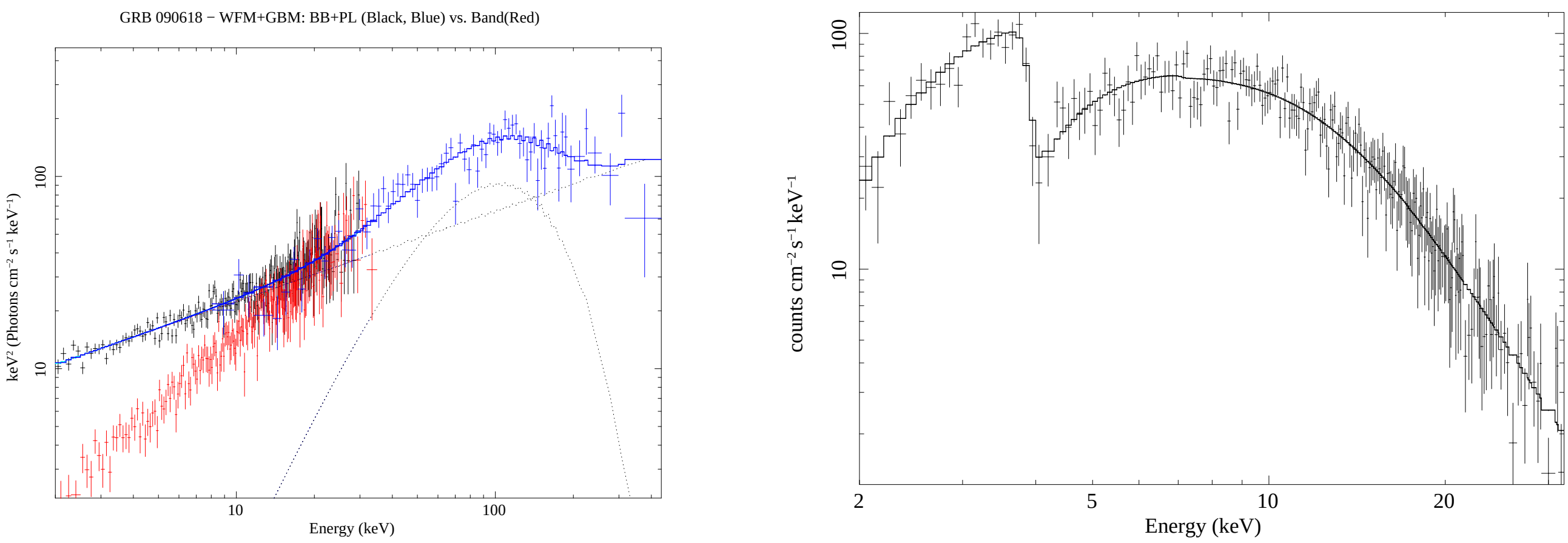

Figure 4: Examples of *LOFT*/WFM simulations of the GRB prompt emission (burst) spectra from. *Left:* Simulated WFM spectra of the first ~50s of GRB 090618 (from Izzo et al. 2012) obtained by assuming either the Band function (black) or the power-law plus black-body model (red) which equally fit the Fermi/GBM measured spectrum, which is also shown (blue). The black–body plus power-law model components best–fitting the Fermi/GBM spectrum are also shown (black dashed lines). *Right:* Simulation of the transient absorption feature in the X-ray energy band detected by BeppoSAX/WFC in the first 8 s of GRB 990705 as would be measured by the *LOFT*/WFM (Amati et al., 2000).

More specifically, the WFM low energy threshold is fundamental for the detection and study of XRFs , high-$z$ GRBs, and soft transient X-ray absorption features, for which energy resolution of 300 eV at energies <10 keV is mandatory. A FOV of about 4 sr combined with a sensitivity of ~0.5 Crab (in 1 s integration time) in the 2–50 keV energy band is required to detect and localize ~120 GRBs per year (including ~1–2 events at $z > 6$, Fig. 3) and perform time resolved spectroscopy of ~2/3 of them.

As estimated with the *BeppoSAX*/WFC and *HETE-2*/WXM $\log N$-$\log S$ and $E_\mathrm{p}$ distributions (Vetere et al., 2007; Sakamoto et al., 2005), about 2/3 of GRBs are X-Ray Flashes (XRFs) or X-Ray Rich (XRRs) GRBs. *LOFT*/WFM has a much better efficiency for the detection of XRFs with respect to main GRB detectors like *CGRO*/BATSE, *Swift*/BAT and *Fermi*/GBM. In Figure 3 (left) we show the expected peak flux sensitivity as a function of the spectral peak energy $E_\mathrm{p}$ (following the method proposed by Band, 2003).

Starting from the $\log N$-$\log P$ from 2–25 keV *HETE-2*/WXM data and: 1) taking into account the FOV difference between *HETE-2*/WXM and *LOFT*/WFM; 2) assuming 80% observation efficiency; 3) assuming a total of 120 *LOFT*/WFM GRB per year, the *LOFT*/WFM will increase the XRF detection per year by a factor of ~7 with respect to *HETE-2*/WXM (Sakamoto et al., 2005). This results in an estimated rate of 36 *LOFT* XRFs per year, or a total number of 108 *LOFT* XRF detections in 3 years (180 for 5 years), to compared for example with the 19 XRF in 3 years detected with *HETE-2*/WXM (Sakamoto et al., 2005).

The extension of the energy band down to ~2 keV of the *LOFT*/WFM will make it more sensitive to very soft XRFs. Even the main GRB detector of the Chinese-French SVOM mission, currently under study, will not be able to explore below 4 keV.

Simulated WFM spectra of the first ~50 s of GRB 090618 were obtained by assuming first a Band function and then a power-law plus black-body model, which describe the *Fermi*/GBM measured spectrum equally well (Izzo et al., 2012). As clearly shown in Fig. 4, the extension of the spectral measurement below 10 keV allowed by the *LOFT*/WFM will make it possible to discriminate between the Band and black-body plus power-law models.

An energy resolution better than 15% (FWHM) for photon energies above 50 keV is enough for accurate measurement of the peak energy of the $\nu F_\nu$ spectrum. A simulation of the transient absorption feature in the X-ray energy band detected in the first 8 s of GRB 990705 is plotted in Fig. 4, together with a simulation of how *LOFT*/WFM would detect the transient blackbody component in the X-ray energy band detected by BeppoSAX/WFC in GRB 990712.

Source location accuracy of a few arcminutes is required to trigger follow-up observations of the detected transients by other telescopes, but it is also essential to achieve the GRB scientific objectives by means of multi-wavelength and multi-messenger studies. For the further advancement of the GRB science it is essential to have in the >2025 time-frame a space instrument capable to continue providing alert service, presently provided by the Swift satellite, to space and ground multi-wavelength telescopes. On board data handling is required to allow prompt follow-up observations (see Sect. 5).

## 4 Perspective GRB Science with the LAD

While the GRB science with the WFM will come 'for free', the contribution from the LAD consisting possibly of the spectral and timing characterization of X-ray afterglow emission up to ~30 keV, will critically depend on the follow-up capabilities and policies of the mission, and it will be strongly dependent on the time that will be required to start a ToO observation.

GRB science that could be performed by the LAD includes (see also Amati et al., Exp. Astr., submitted) the X-ray afterglow spectral and temporal analysis, and the search for emission/absorption lines, predicted by theoretical models if the circum-burst environment were highly metal-enriched. Spectral emission and absorption lines were detected in the X-ray afterglows of only few GRBs, namely GRB 000214 by BeppoSAX (Antonelli et al., 2000), of GRB 991216 by *Chandra* (Piro et al., 2000) and more recently of GRB 130925A by *NuSTAR* (Bellm et al., 2014). However, no such features were detected so far by *Swift*/XRT (Hurkett et al., 2008). Further observations are strongly desirable in order to confirm their existence and ultimately exploit all their potential information on the GRB progenitor, surrounding environment and radiation mechanisms.

Simulations have shown that *LOFT*/LAD capabilities will enable to detect emission lines with characteristics similar to those observed in the past. For example, Figure 6 shows simulations of an emission line at 4.7 keV as the one detected by BeppoSAX/NFI 12 hours after the trigger of GRB 000214. The continuum flux used in the simulation was fixed a factor 2.6 higher with respect to the 2-10 keV flux measured for GRB000214, i.e. at $2 \times 10^{-12}$ erg cm$^{-2}$ s$^{-1}$, corresponding to the expected flux at 6 hours after the trigger (instead of 12 hours), by assuming the measured flux decay index of 1.4 (Antonelli et al. 2000), while the line flux was left unaltered. The line is detected with high statistical confidence.

Providing fast follow-up, e.g., within 5–10 hours from the burst trigger, a significant fraction of GRB X-ray afterglows (~20%) have fluxes above the estimated sensitivity limit for *LOFT*/LAD (Fig. 5). Simulations show that within about 10 hours of trigger, X-ray afterglow fluxes will be detected by *LOFT*/LAD with more efficiency and less integration time than for Swift/XRT. X-ray afterglows to be followed with LAD can be selected among the brightest GRBs, from the burst fluence that correlates with X-ray afterglow fluxes (Gehrels et al., 2008). For these GRBs, *LOFT*/LAD will enable investigations of the plateau phase and its transition to the "normal decay" , which may mark the energy injection by a magnetized neutron star (Nousek et al., 2006; Dall'Osso et al., 2011).

The LAD will also be sensitive to prompt GRB emission, even when the GRB is outside the field of view, that is able to penetrate the LAD collimator. Preliminary Monte Carlo simulations show that, by combining the spectral efficiency of the silicon detectors with the LAD collimator transparency, a collecting area of a few hundred cm$^2$ will be available, where the exact value depends on the off-axis angle and the photon energy (see Marisaldi et al., *LOFT* WP on TGFs). As a consequence, GRBs can be detected through the LAD collimator, enabling us to extend the energy band of the prompt emission up to 80 keV and to perform timing analysis in the 2–80 keV energy band.

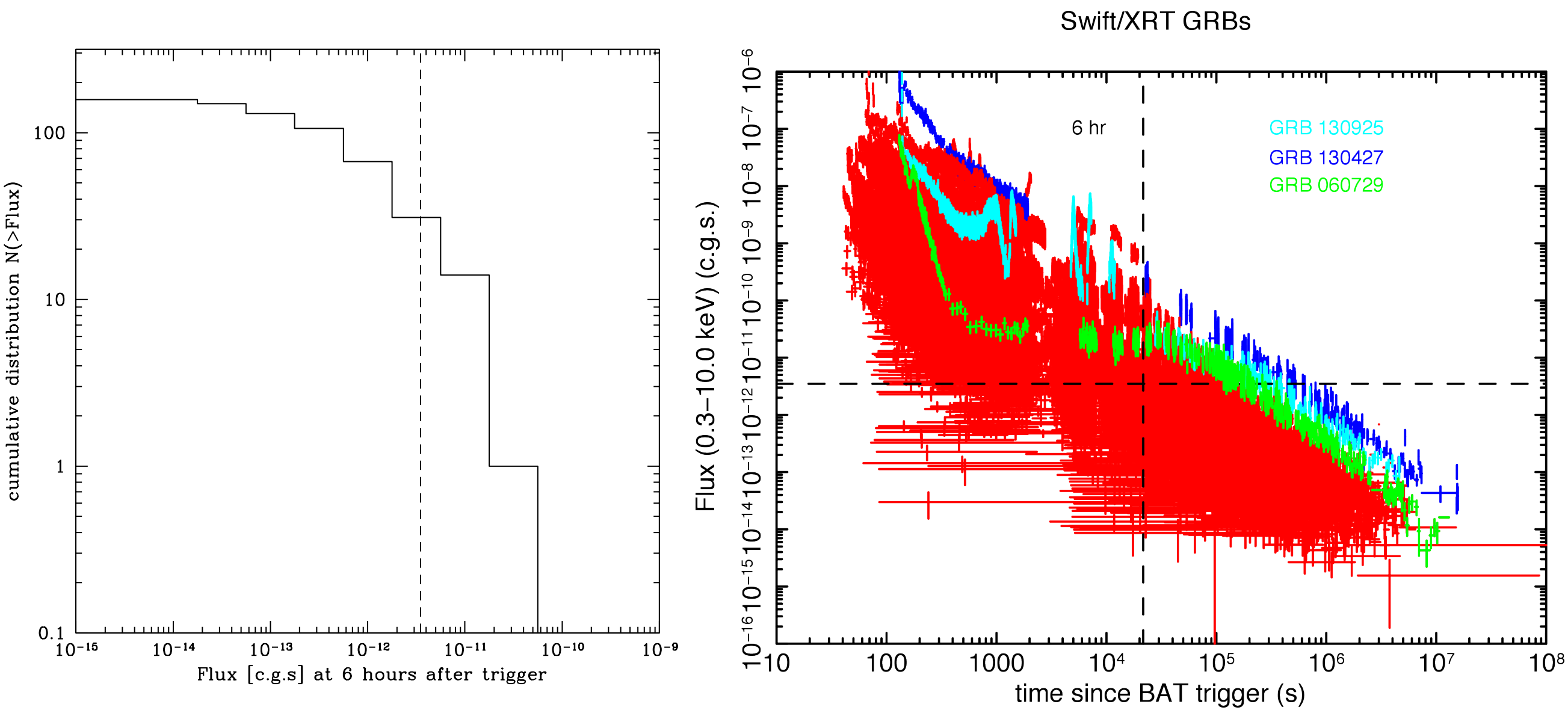

Figure 5: Examples of *LOFT*/LAD capabilities on X-ray afterglow GRB science. *Left:* Cumulative distribution of the X-ray afterglow 0.3–10 keV flux at 6 hours from the GRB onset, as measured by *Swift*/XRT (as of end 2012). The vertical dashed line corresponds to the expected *LOFT*/LAD sensitivity limit (100 s, $5\sigma$, 0.15% accuracy in background estimation). *Right:* X-ray afterglow light curves in 0.3–10 keV measured by *Swift*/XRT as of the end of 2012. The vertical line corresponds to 6 hours from the GRB onset; the horizontal line corresponds to the *LOFT*/LAD sensitivity limit.

## 5 Multiwavelength and multimessenger impact

The WFM will include a near real-time detection capability for bright-events, the so-called *LOFT* Burst Alert System (LBAS, Schanne et al. 2014). The full WFM data stream, which cannot be transmitted to ground in real-time due to the *LOFT* low Earth orbit, is analyzed onboard by the *LOFT* Burst Onboard Trigger (LOBT) in order to detect the appearance of a GRB and localize it on the sky. An onboard VHF transmitter will broadcast the position and trigger time of the brightest events detected by the WFM to a ground system of VHF receivers within about 20–30 seconds. These will cover the equatorial region around the Earth and will guarantee a full coverage of the *LOFT* orbit. Upon reception of an alert, the VHF ground antennas will transmit the trigger times and positions communicated by the LBAS to the *LOFT* Burst Alert Center (LAC) and possibly to ground and space based robotic facilities to ensure quick head-up and proper fast follow-up observations. Once human verified, the alerts will also be broadcast further to the science community at large through fast communication systems (emails, SMS, ATELs, GCNs). For each of the triggers, the WFM will collect data in photon-by-photon mode. These data will be available on the ground for further investigation within a few hours from the trigger. The relatively short orbit of *LOFT* (~90 minutes), and the availability of at least two ground stations for telemetry reception, will ensure that about 20% of the satellite orbit is endowed with a near-real time transfer capability. This guarantees that, for 20% of the triggered events, the corresponding event-by-event data will be available for "immediate" inspection and further broadcasting to the science community at large.

Through its ability to quickly provide GRB triggers and localizations, *LOFT* will

- continue the “service” work for the astrophysical community, carried out presently by Swift of providing detection, prompt localization and temporal/spectral characterization of GRBs, thus allowing their follow-up and multi-wavelength studies with the best telescopes operating post 2020 – the time-domain era – (e.g., *LSST*, *E-ELT*, *SKA*, *CTA*, *Spectrum-X-Gamma*/eROSITA, maybe *XMM-Newton*, *Chandra*, etc.);
- complement simultaneous observations by GRB experiments flying on other satellites by providing low

Figure 6: LAD simulation of the iron emission line at 4.7 keV detected by BeppoSAX/NFI 12 hours after the trigger of GRB 000214, as would be measured by the LAD (Antonelli et al. 2000). The line intensity is $9 \times 10^{-6}$ ph cm$^{-2}$ s$^{-1}$ superposed to an absorbed power law continuum with photon index of $\Gamma = 2.9$, equivalent hydrogen column density of $5.5 \times 10^{20}$ cm$^{-2}$ and a 2-10 keV flux of $2 \times 10^{-12}$ erg cm$^{-2}$ s$^{-1}$.

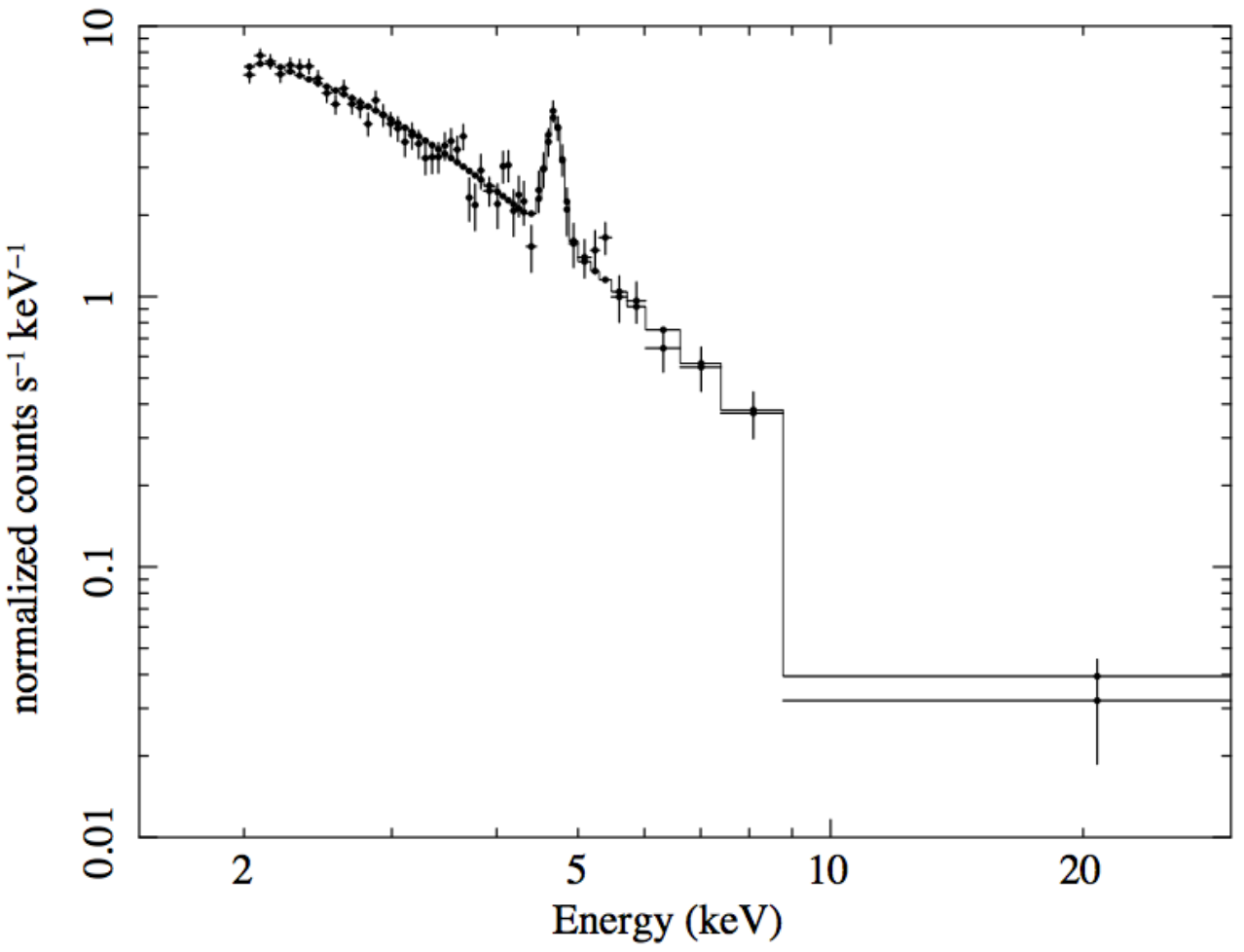

energy extension and GRB position, similarly to what is presently done, for example, by joint analysis of data provided by *Swift*, *Fermi*, and *Konus*/WIND.

The experience gained with GRBs has stressed the importance of quick alerts, reaching the ground within seconds. These alerts have permitted observations that have revealed the brightness of the early afterglow and the rich information transported by photons during this phase. The unique knowledge gained during this phase has proven crucial to constrain the nature of GRB jets, the physical processes at work in the jet, the environment of the source, and will possibly shed new light on the origin of XRFs (see §2).

GRBs are among the best candidate sources of both gravitational waves and of ultra-high energy neutrinos. A prompt follow-up will enable simultaneous observations in the electromagnetic window, making *LOFT* observations a crucial tool for the full exploitation of the GRB role in the upcoming multi-messenger astronomy era. At the time *LOFT* will operate, gravitational waves (GW) are expected to be detected routinely by GW detectors with small error regions, greatly enhancing follow-up, and next generation neutrino detectors are assumed to be operative (e.g. IceCube Gen2). The possibility to detect simultaneously gravitational, neutrinos and electromagnetic waves from the same source is crucial for two main reasons: 1) to enhance the GW and/or neutrino detection significance and 2) to provide a wealth of information on the nature of the source itself.

GRBs, expecially the short GRB class, are among the best candidate sources of gravitational waves and the possibility to detect them through *LOFT*/WFM sky monitoring will provide crucial information on the origin of their progenitors. Short gamma-ray bursts are usually defined as GRBs with burst duration <2 s (Kouveliotou et al., 1993) and are thought to originate from a coalescing binary-system of compact objects. A significant fraction of SGRBs (25% of Swift/BAT SGRBs) are accompanied by longer duration (up to $\sim$ 10 s) extended emission (EE) in the X-ray band (Fong et al., 2013), the origin of which is still a mystery. The low X-ray WFM energy threshold, together with its sensitivity (0.5 mCrab in 1 sec), would enable the detection of the short-GRBs extended emission up to 50–100 s after the trigger for short GRB with X-ray fluxes similar to GRB 050724, 060614, 061006, 061210, 080503, that are taken as an example among the 14 short GRBs with EE emission detected by Swift (Fong et al., 2013; Gompertz et al., 2013). It is possible that such decrease in the detector energy threshold may dramatically increase the number of SGRB EE observations performed so far (even after SVOM era) and thus increase the number of simultaneous detections of GW.

## 6 Conclusions

No past, present, or future GRB experiment has such a combination of imaging, low energy threshold and high energy resolution (and wide FOV), which will make the *LOFT*/WFM unique for GRB science. For instance, the *BeppoSAX*/WFC or the *HETE-2*/WXM had a low energy threshold of around 2 keV, but with much worse energy resolution and smaller field of view. *SVOM* (planned to be active from 2021) will have a low energy threshold of 4 keV, but with significantly worse energy resolution and a smaller FOV with respect to the *LOFT*/WFM.

Because of their complex phenomenology and extreme energetics, the study of GRBs is relevant to several different fields of astrophysics, ranging from plasma and black-hole physics to cosmology. In particular, the *LOFT* measurements described above will provide a major step forward in the understanding of GRB physics, progenitors and their sub-classes. *LOFT*, in combination with multi-wavelength observations, will also shed light on fundamental topics like the evolution of the star forming rate, the galactic ISM up to the epoch of re-ionization, the first generation (population III) stars, the understanding of the diversity and rate of core-collapse SNe, the measurement of cosmological parameters.

In addition, GRB science already involves observers and instrumentation from different communities: optical/IR robotic telescopes (prompt detections and localization of the optical counterparts), major optical/IR telescopes such as the VLT, Gemini, Hubble, JWST (identification, redshift and ISM of the host galaxies, optical afterglow decay and jet signatures), radio telescopes like VLA (afterglow modelling and energetics) observations in the TeV range by Cherenkov telescopes (challenging emission models). By 2025, gravitational wave (GW) detectors will routinely detect GW sources among which GRBs are the most promising candidates, making *LOFT* an active participant to the GW Astronomy era.